\documentclass[apjl]{emulateapj}
\usepackage{apjfonts,amssymb}
\usepackage{amsbsy}

\def\apj{ApJ}
\def\apjs{ApJS}
 
\def\aj{AJ}
\def\apjl{ApJL}
\def\jfm{J. Fluid Mech.}
\def\jgr{J. Geophys. Research}

\def\pha{Physica A}
\def\phf{Phys. Fluids}
\def\pre{Phys. Rev. E}
\def\prl{Phys. Rev. Lett.}
\def\aap{A\&A}
\def\araa{ARA\&A}
\def\mnras{MNRAS}
\def\jcp{J. Comput. Phys.}
\slugcomment{ApJ 665, 000--000, 2007 August 20}
\shorttitle{SIMULATIONS OF SUPERSONIC TURBULENCE}
\shortauthors{KRITSUK, NORMAN, PADOAN, \& WAGNER}

\begin{document}
\title{The Statistics of Supersonic Isothermal Turbulence}

\author{Alexei G. Kritsuk\altaffilmark{1}, Michael L. Norman, Paolo Padoan, and Rick Wagner}
\affil{Department of Physics and Center for Astrophysics and Space Sciences,
University of California, San Diego,\\
9500 Gilman Drive, La Jolla, CA 92093-0424;
akritsuk@ucsd.edu,
mlnorman@ucsd.edu,
ppadoan@ucsd.edu
}

\altaffiltext{1}{Also at: Sobolev Astronomical Institute,
St. Petersburg State University, St. Petersburg, Russia.}

\begin{abstract} 
We present results of large-scale three-dimensional simulations of supersonic
Euler turbulence with the piecewise parabolic method and multiple grid 
resolutions up to $2048^3$ points. 
Our numerical experiments describe non-magnetized driven turbulent flows 
with an isothermal equation of state and an rms Mach number of~6. We discuss 
numerical resolution issues and demonstrate convergence, in a statistical sense,  
of the inertial range dynamics in simulations on grids larger than $512^3$ points. 
The simulations allowed us to measure the absolute velocity scaling exponents
for the first time.
The inertial range velocity scaling in this strongly compressible regime  deviates 
substantially from the incompressible Kolmogorov laws. The slope of the velocity 
power spectrum, for instance, is~$-1.95$ compared to~$-5/3$ in the 
incompressible case. The exponent of the third-order velocity structure function 
is~$1.28$, while in incompressible turbulence it is known to be unity. 
We propose a natural extension of Kolmogorov's phenomenology that takes into account 
compressibility by mixing the velocity and density statistics and preserves the 
Kolmogorov scaling of the power spectrum and structure functions of the 
density-weighted velocity $\pmb{v}\equiv\rho^{1/3}\pmb{u}$. The low-order
statistics of $\pmb{v}$ appear to be invariant with respect to changes in the Mach 
number. For instance, at Mach 6 the slope of the power spectrum of $\pmb{v}$ is~$-1.69$, and
the exponent of the third-order structure function of $\pmb{v}$ is unity. 
We also directly measure the mass dimension of the ``fractal'' density 
distribution in the inertial subrange, $D_m\approx 2.4$, which is similar to
the observed fractal dimension of molecular clouds and agrees well with the 
cascade phenomenology. 
\end{abstract}

\keywords{
hydrodynamics ---
instabilities ---
ISM: structure ---
methods: numerical --- 
turbulence
}

\section{Introduction}
Understanding the nature of supersonic turbulence is of fundamental importance
in both astrophysics and aeronautical engineering. In the interstellar medium (ISM), 
highly compressible turbulence is believed to control star formation in dense 
molecular clouds \citep{padoan.02}.
In radiation-driven outflows from carbon-dominant Wolf-Rayet stars,
supersonic turbulence creates highly clumpy structure that is stochastically 
variable on a very short time-scale \citep[e.g.,][]{acker...02}.
Finally, a whole class of more {\em terrestrial} applications deals 
with the drag and stability of projectiles traveling through the air at 
hypersonic speeds.

Molecular clouds have an extremely inhomogeneous structure and the intensity of 
their internal motions corresponds to an rms Mach number of order 20. 
\citet{larson81} has demonstrated that within the range of scales from 0.1
to 100~pc, the gas density and the velocity dispersion tightly correlate with 
the cloud size.\footnote{For an earlier version of what
is now known as Larson's relations, see \citet{kaplan.53} 
and also see \citet{brunt03} for the latest results on the velocity 
dispersion--cloud size relation.} Supported by other 
independent observational facts indicating 
scale invariance, these relationships are often interpreted in terms of 
supersonic turbulence with characteristic Reynolds numbers $Re\!\sim\!10^8$
\citep{elmegreen.04}. 
Within a wide range of densities above $10^3$~cm$^{-3}$, the gas temperature
remains close to $\!\approx\!10$~K, since the thermal equilibration time at these
densities is shorter than a typical hydrodynamic (HD) timescale. Thus, an isothermal 
equation of state can be used as a reasonable approximation. Self-gravity, 
magnetic fields, chemistry, cooling, and heating, as well as radiative transfer, 
should ultimately be accounted for in turbulent models of molecular clouds.
However, since highly compressible turbulence still remains an unsolved problem
even in the absence of magnetic fields, our main focus here is specifically on 
the more tractable HD aspects of the problem. 

Magnetic fields are important for the general ISM dynamics and, particularly, for 
the star formation process. Observations of molecular clouds are consistent 
with the presence of super-Alfv\'enic turbulence \citep{padoan...04a}, and thus, the 
average magnetic field strength may be much smaller than required to support the 
clouds against the gravitational collapse \citep{padoan.97,padoan.99}. Even weak 
fields, however, have the potential to modify the properties of supersonic turbulent 
flows through the effects of magnetic tension and magnetic pressure and introduce
small-scale anisotropies.
Magnetic tension tends to stabilize hydrodynamically unstable post-shock 
shear layers \citep{miura.82,keppens...99,ryu..00,baty..03}. The shock jump 
conditions modified by magnetic pressure result in substantially different 
predictions for the initial mass function of stars forming in non-magnetic and 
magnetized cases via turbulent fragmentation \citep{padoan....07}.  Remarkably, 
although not surprisingly \citep[e.g.,][]{armi.85}, the impact
of magnetic field on the low-order statistics of super-Alfv\'enic turbulence appears
to be rather limited. At a grid resolution of $1024^3$ points, the slopes of the 
power spectra in our simulations show stronger sensitivity to the numerical 
diffusivity of the scheme of choice than to the presence of the magnetic field 
\citep{padoan....07}. 
The similarity of non-magnetized and weakly 
magnetized turbulence will allow us to compare our HD results with those from 
super-Alfv\'enic magnetohydrodynamic (MHD) simulations where equivalent pure 
HD simulations are not available in the literature.
\begin{figure}
\epsscale{1.15}
\centering
\plotone{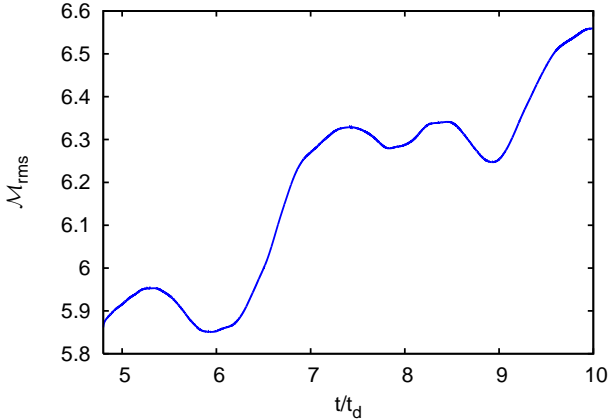}
\caption{Time evolution of the rms Mach number in the $1024^3$ simulation of driven 
Mach~6 turbulence.}
\label{mach}
\end{figure}

Numerical simulations of {\em decaying} supersonic HD turbulence with the piecewise
parabolic method (PPM) in two dimensions were pioneered by 
\citet{passot..88}\footnote{See a review on 
compressible turbulence by \citet{pouquet..91} for references to earlier works.} and 
then followed up with high-resolution two- and three-dimensional 
simulations by \citet{porter..92a,porter..92b,porter..94,porter..98}. 
\citet{sytine....00} compared the results of PPM Euler computations 
with PPM Navier-Stokes results and showed that the Euler simulations agree well with 
the high-Re limit attained in the Navier-Stokes models. The convergence
in a statistical sense as well as the direct comparison of structures in configuration
space indicate the ability of PPM to accurately simulate turbulent flows over a wide
range of scales.
More recently, \citet{porter..02} discussed measures of intermittency in simulated 
{\em driven} transonic flows at Mach numbers of the order unity on grids of up to 
$512^3$ points.
\citet{porter...99} review the results of these numerical studies, focusing on
the origin and evolution of turbulent structures in physical space as well as on
scaling laws for two-point structure functions. One of the important results
of this fundamental work is the demonstration of the compatibility of a Kolmogorov-type
\citep[K41]{kolmogorov41a} spectrum with a {\em mild} gas compressibility at 
transonic Mach numbers.

Since most of the computations discussed above assume a perfect gas equation
of state with the ratios of specific heats $\gamma=7/5$ or $5/3$ and Mach 
numbers generally below 2, the question remains whether this result will still hold
for near isothermal conditions and {\em hypersonic} Mach numbers characteristic of dense 
parts of star-forming molecular clouds where the gas compressibility is much higher.
What kind of coherent structures should one expect to see in highly supersonic 
turbulence? Do low-order statistics of turbulence follow the K41 predictions closely 
in this regime? 
How can the statistical diagnostics traditionally used in studies of incompressible
turbulence be extended to reconstruct the energy cascade properties in supersonic flows?
How can we measure the intrinsic intermittency of supersonic turbulence? 
Many of these and similar questions can only be addressed with numerical simulations 
of sufficiently high resolution.
The interpretation of astronomical data from new surveys of the cold ISM and dust in
the Milky Way by the {\em Spitzer} and {\em Herschel Space Observatory} satellites 
requires more detailed knowledge of these basic properties of supersonic turbulence.

In this paper we report the results from large-scale numerical simulations
of driven supersonic isothermal turbulence at Mach 6 with PPM and grid resolutions 
up to $2048^3$ points. The paper is organized as follows: Section \ref{mthd} 
contains the details of the simulations' setup and describes the input parameters. 
The statistical diagnostics, including power spectra of the velocity, the kinetic 
energy and the density, and velocity structure functions, together with a discussion 
of turbulent structures and their fractal dimension are presented in Section 
\ref{rslt}. In Section \ref{mdl} we combine the scaling laws determined in our 
numerical experiments to verify a simple compressible cascade model proposed by 
\citet{fleck96}. We also introduce a new variable $\pmb{v}\equiv\rho^{1/3}\pmb{u}$ 
that controls the energy transfer rate through the compressible cascade. We then 
summarize the results and discuss possible ways to validate our numerical model with 
astrophysical observations in Sections \ref{dscs} and \ref{cncl}.

\section{Methodology} \label{mthd}
We use PPM implemented in the {\sl Enzo} code\footnote{See http://lca.ucsd.edu/} to solve the 
Euler equations for the gas density $\rho$ and the velocity 
$\pmb{u}$ with a constant external acceleration term $\pmb{F(x)}$ such that $\left<\pmb{F(x)}\right>=0$,
\begin{equation}
\partial_t\rho+\pmb{\nabla}\cdot(\rho\; \pmb{u})=0,
\label{1}
\end{equation}
\begin{equation}
\partial_t\pmb{u}+\pmb{u}\cdot\pmb{\nabla}\pmb{u}=-\pmb{\nabla}\rho/\rho+\pmb{F},
\label{2}
\end{equation}
in a periodic box of linear size $L=1$ starting with an initially uniform density 
\begin{figure}
\epsscale{1.15}
\centering
\plotone{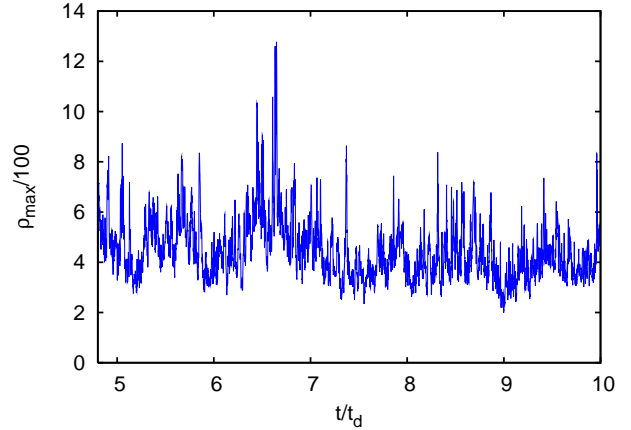}
\caption{Same as Fig. \ref{mach}, but for
the maximum gas density.}
\label{dmax}
\vspace{3mm}
\end{figure}
distribution $\rho(\pmb{x},t=0)=\rho_0(\pmb{x})\equiv1$ and 
assuming the sound speed $c\equiv1$. The equations that were actually numerically integrated
were written in a conventional form of conservation laws for the mass, momentum, and 
total energy that is less compact, but {\em nearly} equivalent to equations (\ref{1}) and (\ref{2}), 
since in practice we mimic the isothermal equation of state by setting the specific heats 
ratio in the ideal gas equation of state very close to unity, $\gamma=1.001$.

The simulations were initialized on grids of $256^3$ or  $512^3$ points with a random 
velocity field $\pmb{u}(\pmb{x},t=0)=\pmb{u}_0(\pmb{x})\propto\pmb{F}$ that contains 
only large-scale power within the range of 
wavenumbers $k/k_{min}\in[1,2]$, where $k_{min}=2\pi$, and that corresponds to the rms Mach 
number ${\cal M}=6$. The dynamical time is hereafter defined as $t_d\equiv L/(2{\cal M})$.

\subsection{Uniform Grid Simulation at $1024^3$} 
Our major production run is performed on a grid of $1024^3$ points that allowed us to resolve
a portion of the uncontaminated inertial range sufficient to get a first approximation for
the low-order scaling exponents.

We started the simulation at a lower resolution of
$512^3$ points and evolved the flow from the initial conditions for five dynamical times to 
stir up the gas in the box.
Then we doubled the resolution and evolved the simulation for another $5t_d$ on a grid of 
$1024^3$ points.

The time-average statistics were computed
using 170 snapshots evenly spaced in time over the final segment of $4\;t_d$.
(We allowed one dynamical time for flow relaxation at high resolution,
so that it could reach a statistical steady state after regridding.)
We used the full set of 170 snapshots to derive the density statistics, since the density field
displays a very high degree of intermittency. This gave us a very large statistical sample,
e.g. $2\times 10^{11}$ measurements were available to determine the probability density function 
(PDF) of the gas density discussed in Section \ref{lognorm}. 
The time-average power spectra discussed in Sections \ref{vps}, \ref{eps}, \ref{dps}, and 
\ref{mdl} are also based on the full data set.
The velocity structure functions presented in Section \ref{vsf} are derived from a sample of 20\%
of the snapshots covering the same period of $4\;t_d$. The corresponding two-point PDFs 
of velocity differences were built on $2-4\times10^9$ pairs per snapshot each, depending on 
the pair separation.

\subsection{AMR Simulations}
The adaptive mesh refinement (AMR) simulation with effective resolution of $2048^3$ 
points was initialized by
evolving the flow on the root grid of $512^3$ points for six dynamical times. Then
one level of refinement by a factor of 4 was added that covers on average 50\% 
of the domain volume. The grid is refined to better resolve strong shocks and to capture 
HD instabilities in the layers of strong shear.
We use the native shock-detection algorithm of PPM, and we flag for refinement those zones that 
are associated with shocks with density jumps in excess of 2. We also track shear layers using
the Frobenius norm $\|\partial_i u_j(1 - \delta_{ij})\|_F$, see also \citet{kritsuk..06}. This
second refinement criterion adds about 20\% more flagged zones that would be left unrefined 
if only the refinement on shocks were used. The simulation was continued with AMR for only 
$1.2\: t_d$, which allows enough time for relaxation of the flow at high resolution, 
but does not allow us to perform time-averaging over many statistically independent snapshots. 
Therefore, we use the data from AMR simulations mostly to compare the quality of 
instantaneous statistical quantities from simulations with adaptive and nonadaptive meshes.

\subsection{Random Forcing \label{force}}
The initial velocity field is used, after an appropriate renormalization, as a steady 
random force (acceleration) to keep the total kinetic energy within the box on an 
approximately constant level during the simulations. The force we applied is isotropic 
in terms of the total specific kinetic energy per dimension, 
$\overline{u_{0,x}^2}=\overline{u_{0,y}^2}=\overline{u_{0,z}^2}$,
but its solenoidal ($\pmb{\nabla}\cdot\pmb{u}_{0,S}\equiv0$) and
dilatational ($\pmb{\nabla}\times\pmb{u}_{0,D}\equiv0$)
components are anisotropic, since one of the three directions is dominated by the large-scale
compressional modes, while the other two are mostly solenoidal. The distribution of the total
specific kinetic energy 
\begin{equation}
{\cal E}\equiv\frac{1}{2}\int_{\cal V} u^2dV={\cal E}_S+{\cal E}_D
\end{equation}
between the solenoidal ${\cal E}_S$ and dilatational ${\cal E}_D$ components is such that 
$\chi_0\equiv {\cal E}_{0,S}/{\cal E}_0\approx0.6$. The forcing field is helical, but the 
mean helicity is very low, $\overline{h_0}^2\ll\overline{h_0^2}$, where the helicity $h$ 
is defined as 
\begin{equation}
h\equiv\pmb{u}\cdot\pmb{\nabla}\times\pmb{u}.
\end{equation}
Note that in compressible flows with an isothermal equation of state the mean 
helicity is conserved, as in the incompressible case, since the Ertel's potential 
vorticity is identically zero \citep{gaffet85}.

\section{Results}\label{rslt}
In this Section we start with a general quantitative description of our simulated
turbulent flows and then derive their time-average statistical properties. We
end up with assembling the pieces of this statistical picture in a context of a
simple cascade model that extends the Kolmogorov phenomenology of incompressible 
turbulence into the compressible regime.

\subsection{Time-evolution of Global Variables} 
\label{glb}
The time variations of the rms Mach number and of the maximum gas density 
in the $1024^3$ simulation are shown in Figs.~\ref{mach} and~\ref{dmax}. The kinetic energy 
oscillates between 18 and 22, roughly following the Mach number evolution. Note the highly 
intermittent bursts of activity in the plot of $\rho_{max}(t)$.
The time-average enstrophy 
\begin{equation}
\Omega\equiv\frac{1}{2}\int_{\cal V}\left|\pmb{\omega}\right|^2dV\approx10^5
\end{equation}
and the Taylor scale 
\begin{equation}
\lambda\equiv\sqrt{\frac{5E}{\Omega}}\approx0.03=30\Delta, 
\end{equation}
where $\Delta$ is the linear grid spacing and $\pmb{\omega}\equiv\pmb{\nabla}\times\pmb{u}$ 
is the vorticity.
The rms helicity grows by a factor of $7.7$ in the initial phase of the simulation and then
remains roughly constant at a level of $1.2\times10^3$. 
While the conservation of the mean helicity is not built directly into the numerical method,
it is still satisfied reasonably well. The value of $\overline{h}(t)$ is 
contained within $\pm2$\% of its rms value during the whole simulation.

If, instead of the Euler equations, we were to consider the Navier-Stokes equations with the
explicit viscous terms, we could estimate the total viscous dissipation rate within the
computational domain,
\begin{equation}
\epsilon=
\int_{\cal V}\Phi_i u_i\; d V,
\label{13}
\end{equation}
where 
\begin{equation}
\Phi_i\equiv\frac{2}{Re}\frac{\partial}{\partial x_j}\left(\mathfrak{S}_{ij}-
\frac{1}{3}\nabla\cdot u\;\delta_{ij}
\right)
\end{equation}
and
\begin{equation}
\mathfrak{S}_{ij}\equiv\frac{1}{2}\left(\frac{\partial u_i}{\partial x_j}+
\frac{\partial u_j}{\partial x_i}\right)
\end{equation}
is the symmetric rate-of-strain tensor, and the integration is done over the volume of
the domain ${\cal V}=1$.
Using vector identities,  $\pmb{\Phi}$ can be 
rewritten through the vorticity $\pmb{\omega}$ and the dilatation $\pmb{\nabla}\cdot\pmb{u}$ as
\begin{equation}
\pmb{\Phi} =
-\frac{1}{Re}\left(\pmb{\nabla}\times\pmb{\omega}-
\frac{4}{3}\pmb{\nabla}\pmb{\nabla}\cdot\pmb{u}\right),
\end{equation}
and, by partial integration in equation (\ref{13}), 
\begin{equation}
\epsilon=-\frac{1}{Re}
\left(\left<|\pmb{\nabla}\times\pmb{u}|^2\right>
+\frac{4}{3}\left<|\pmb{\nabla}\cdot\pmb{u}|^2\right>
\right),
\label{dss}
\end{equation}
where the two terms on the right-hand side of equation (\ref{dss}) describe the mean dissipation rate due to 
solenoidal and dilatational velocities, respectively, $\epsilon=\epsilon_S + \epsilon_D$.

\begin{figure}
\epsscale{1.15}
\centering
\plotone{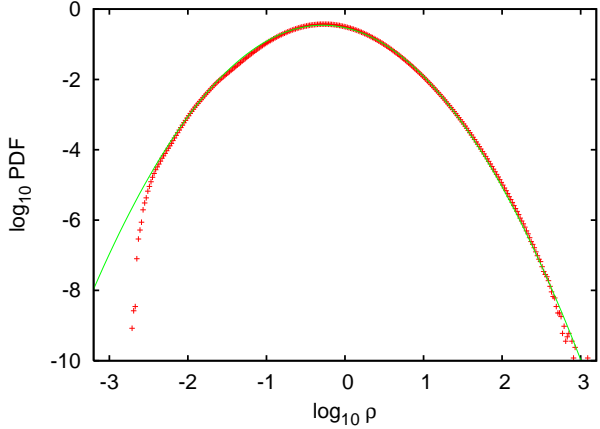}
\caption{Probability density function for the gas density and the best-fit lognormal
approximation. Note the excellent fit quality over eight decades in the probability
along the high-density wing of the PDF. The sample size is $2\times10^{11}$.}
\label{dpdf}
\end{figure}
In our Euler simulations, the role of viscous dissipation is played by the numerical diffusivity
of PPM. This numerical dissipation does not necessarily have the same or similar functional 
form as the one used in the Navier-Stokes equations. It is still instructive, however, to
get a flavor of the numerical dissipation rate by studying the properties of the vorticity 
and dilatation. The so-called small-scale compressive ratio \citep{kida.90,kida.92},
\begin{equation}
r_{CS}\equiv\frac{\left<|\pmb{\nabla}\cdot\pmb{u}|^2\right>}{\left<|\pmb{\nabla}\cdot\pmb{u}|^2\right>+
\left<|\pmb{\nabla}\times\pmb{u}|^2\right>},
\end{equation}
represents the relative importance of the dilatational component at small scales.
The time-average $\overline{r_{CS}}\approx0.28$. The mean fraction of the dissipation rate 
that depends solely on the solenoidal velocity component, $\epsilon_S/\epsilon\approx0.65$. 
Even though the two-dimensional shock fronts are dominating the geometry of the density 
distribution in the dissipation range (see Section \ref{fr}), the dissipation rate 
itself is dominated by the solenoidal velocity component that tracks corrugated 
shocks due to vortex stretching in the associated strong shear flows.

\subsection{Density PDF} 
\label{lognorm}
\begin{figure}
\epsscale{1.15}
\centering
\plotone{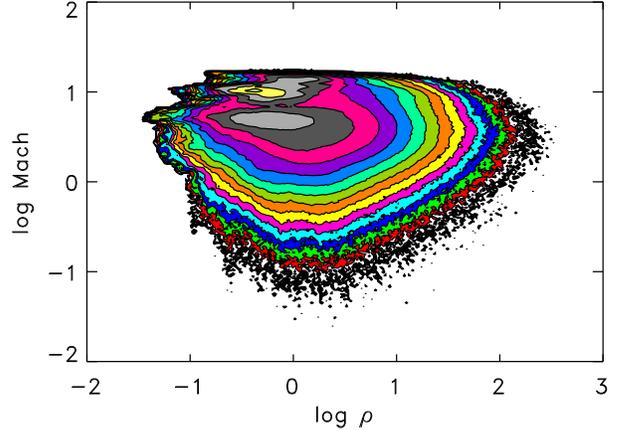}
\caption{Two-dimensional probability density function for the gas density and Mach number.
The data here are based on a subvolume of $700\times500\times250\approx10^8$ points from a
representative snapshot at $t=7\;t_d$ that we will also use later in Figs. \ref{sli} and 
\ref{morph}. The contours show the probability density levels separated by factors of 2.}
\label{dm99}
\vspace{2mm}
\end{figure}

In isothermal turbulence the gas density does not correlate with the
local Mach number. As a result, the density PDF follows a lognormal 
distribution \citep{vazquezsemadeni94,padoan..97,passot.98,nordlund.99,biskamp03}. 
Fig.~\ref{dpdf} shows our results for the time-average density PDF and its best-fit
lognormal representation
\begin{equation}
p(\ln \rho) d \ln \rho = \frac{1}{\sqrt{2\pi\sigma^2}}\times \exp \left[-\frac{1}{2}
\left(\frac{\ln \rho - \overline{\ln \rho}}{\sigma}\right)^2\right] d \ln \rho,
\end{equation}
where the mean of the logarithm of the density, $\overline{\ln \rho}$, is determined by
\begin{equation}
\overline{\ln \rho} = -\sigma^2/2.
\end{equation}
The lack of a Mach number--density correlation is illustrated in
Fig.~\ref{dm99} where we show the two-dimensional PDF for the density and Mach number. 
The density distribution is very broad due to the very high 
degree of compressibility under isothermal supersonic conditions. 
The density contrast $\sim\!10^6$ is orders of magnitude higher than in the transonic 
case at $\gamma=1.4$ studied by \citet{porter..02}. 
Note the excellent quality of the lognormal fit in Fig. \ref{dpdf}, particularly at 
high densities, over more than eight decades in the probability and a very low noise 
level in the data. 

If we express the standard deviation $\sigma$ as a function of Mach number 
${\cal M}$ as
\begin{equation}
\sigma^2=\ln\:(1+b^2{\cal M}^2), 
\end{equation}
we get the best-fit value of $b\approx0.260\pm0.001$ for $\log_{10}\rho\in[-2,2]$, 
which is smaller than the $b\approx0.5$ determined in \citet{padoan..97} for supersonic 
MHD turbulence. The powerful intermittent bursts in $\rho_{max}(t)$ (see 
Fig.~\ref{dpdf}) correspond to large departures from the time-average PDF caused 
by head-on collisions of strong shocks. These events are
usually followed by strong rarefactions that are seen as large oscillations in the 
low-density wing of the PDF and also in the density power spectrum. Intermittency is 
apparently very strong in supersonic turbulence. 

\subsection{Velocity Power Spectra, Bottleneck Phenomenon, Numerical Dissipation, 
and Convergence\label{vps}} 
We define the velocity power spectrum ${\cal E}(\pmb{k})$ (or the {\em specific} 
kinetic energy spectral density) in terms of the Fourier transform of the velocity 
$\pmb{u}$
\begin{equation}
\pmb{\tilde{u}}(\pmb{k})=
\frac{1}{(2\pi)^3}\int_{\cal V}\pmb{u}(\pmb{x})e^{-2\pi i \,\pmb{k}\cdot\pmb{x}}d\pmb{x}
\end{equation}
as the square of the Fourier coefficient
\begin{equation}
{\cal E}(\pmb{k})\equiv\frac{1}{2}\left|\pmb{\tilde{u}}(\pmb{k})\right|^2,
\end{equation}
and then we define the three-dimensional velocity power spectrum as
\begin{equation}
{\cal E}(k)\equiv\int_{\tilde{\cal V}} {\cal E}(\pmb{k})\delta(|\pmb{k}|-k)d\pmb{k},
\end{equation}
where $\delta(k)$ is the Dirac $\delta$-function. The integration of the velocity power 
spectrum gives the specific kinetic energy (cf. eq. [3]),
\begin{equation}
{\cal E}\equiv\int {\cal E}(k) d k = \frac{1}{2}\left<\pmb{u}^2\right>.
\end{equation}

Our main focus in this Section is on the self-similar scaling of the power spectrum 
in the inertial range
\begin{equation}
{\cal E}(k)\sim k^{-\beta}
\end{equation}
that is limited by the kinetic energy input from the random force on large scales 
($k/k_{min}< 2$) and by the spectrum flattening in the near-dissipation part 
of the inertial range due to the so-called bottleneck effect related to a 
three-dimensional non-local mechanism of energy transfer between modes of differing 
length scales \citep{falkovich94}. 

\begin{figure}
\epsscale{1.15}
\centering
\plotone{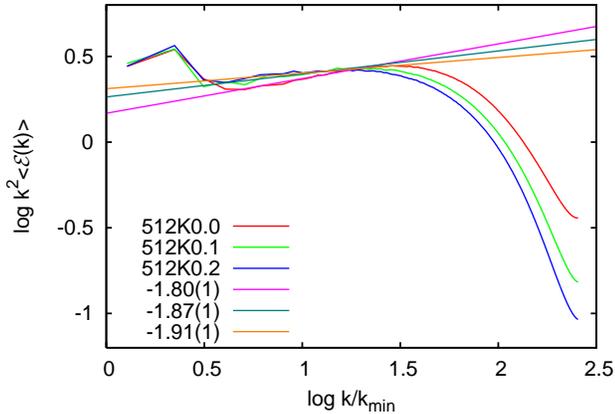}
\caption{Time-averaged velocity power spectra as a function of PPM diffusion coefficient
$K=0.0,\; 0.1$, and $0.2$ at resolution of $512^3$ grid points. The straight lines show the best
linear fits to the spectra obtained for $\log k/k_{min} \in [0.6, 1.3]$. The inertial range is barely
resolved, since the driving scale overlaps with the bottleneck-contaminated interval. 
The slope of the ``flat'' part of the spectrum is mainly controlled by numerical diffusion.}
\label{conv512}
\vspace{3mm}
\end{figure}

The bottleneck phenomenon has been observed both experimentally and in numerical simulations 
\citep[e.g.,][]{porter..94,kaneda....03,dobler...03,haugen.04}.
The strength of the bottleneck depends on the way the dissipation scales with the wavenumber, 
and the effect is more pronounced when the dissipation grows faster than $\sim k^2$
\citep{falkovich94}. When we numerically integrate the Euler equations using PPM, 
the dissipation is solely determined by the method and affects scales smaller 
than $32\Delta$ \citep{porter.94}. According to 
\citet{porter..92a}, the effective numerical viscosity of PPM has a wavenumber 
dependency intermediate between $\sim k^4$ and $\sim k^5$.
The high-order basic reconstruction scheme of PPM is designed to improve 
resolution in shocks and contact discontinuities and, therefore, numerical diffusion is
controlled by various switches and is nonuniform in space. 

An addition of small diffusive flux near shocks\footnote{This is controlled by the 
parameter $K$ in equation (4.5) in \citet{colella.84}.} changes the scaling properties of 
numerical dissipation and reduces the flattening of the spectrum due to the bottleneck 
phenomenon. When the grid is not large enough to resolve the basic flow, the bottleneck 
bump in the spectrum is smeared and can be easily misinterpreted as a shallower
spectrum. Figure~\ref{conv512} shows how the slope of the ``flat'' section in the
velocity power spectrum at $512^3$ depends on the value of the diffusion coefficient $K$.
The inertial range is unresolved in all three cases shown. The measured slope 
$\beta(K)$ changes from $1.91\pm0.01$ to $1.80\pm0.01$ as the diffusion
coefficient decreases from 0.2 to zero and the bottleneck bump gets stronger. While the
statistical uncertainty of estimated power indices based on the fitting procedure is 
rather small, the variation of $\beta$ as a function of the diffusion coefficient is
substantial and accounts for $\sim33$\% of the difference between Burgers and 
Kolmogorov slopes! 
Clearly, it is difficult to get reliable estimates for the inertial range power indices from
simulations with resolution up to $512^3$, since the slope of the spectrum is so dependent 
on numerical diffusivity. To study higher Mach number flows with PPM, an even higher 
resolution would be required. Note that with other numerical methods, 
one typically needs to use even larger grids, since the amount of dissipation provided by 
PPM is quite small compared to what is usually given in finite-difference schemes.
In low-resolution simulations with high numerical dissipation that depends on the 
wavenumber as $\sim k^2$ or weaker, the power spectra appear steeper than 
they would have been when properly resolved.
\begin{figure}
\epsscale{1.15}
\centering
\plotone{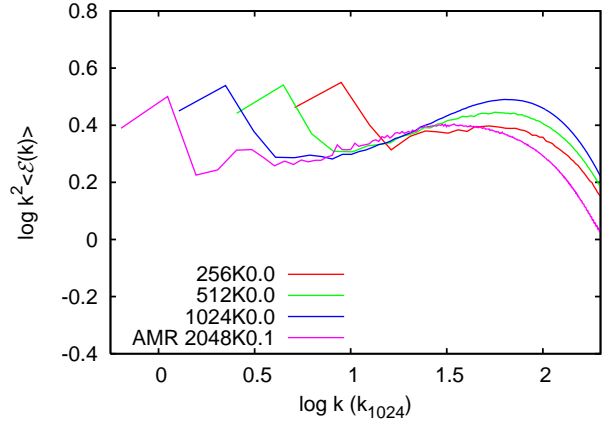}
\caption{Compensated velocity power spectra from uniform grid PPM simulations at resolutions 
$256^3$, $512^3$, $1024^3$, and from AMR simulation with effective resolution of $2048^3$
grid points. The wavenumbers are normalized to match $k/k_{min}$ of the $1024^3$ simulations 
at the Nyquist frequency. The diffusion coefficient $K=0$ for the uniform grid simulations, and
$K=0.1$ for the AMR simulation. All power spectra are time-averaged over 
$t\in[6, 10]\;t_d$ with the exception of the AMR one that is taken at $t=7.2\;t_d$. 
The spectra demonstrate convergence for the inertial range of scales.}
\label{conv1024AMR}
\end{figure}

In simulations with AMR, the grid resolution is nonuniform, and
scaledependence (as well as nonuniformity) of numerical diffusion becomes even more complex. 
One should generally expect a more extended range of wavenumbers affected by 
numerical dissipation in AMR simulations than in uniform grid simulations, when the
effective resolution is the same. 
The dependence of the effective numerical diffusivity on the wavenumber in the AMR runs
is, therefore, weaker than in simulations on uniform grids, and the bottleneck bump should 
be suppressed. 

In Figure~\ref{conv1024AMR} we combined the velocity power spectra from the unigrid 
simulations at $256^3$, $512^3$, and $1024^3$ grid points with the diffusion coefficient 
$K=0$ to illustrate the convergence of the inertial range scaling with the improved 
resolution. 
All three simulations were performed with the same large-scale driving force, and the 
spectra were averaged over the same time interval, so the only difference between them
is the resolution-controlled effective Reynolds number. We plotted the compensated
spectra $k^2{\cal E}(k)$ to specifically exaggerate the relatively small changes in
the slopes. The over-plotted power spectrum from a single snapshot from our largest
to date AMR simulation with the effective resolution of $2048^3$ grid points seems to confirm
the (self-) convergence. Thus, one may hope that our estimates of the scaling exponents
based on the time-averaged statistics from the $1024^3$ simulation will not be too far 
off from those that correspond to Re$\;\to\infty$. We further validate this statement
in Section \ref{vsf}, where we discuss the scaling properties of the velocity structure
functions. Note also the suppression of the bottleneck bump visible in the AMR spectrum
that occurs due to the higher effective diffusivity as we discussed earlier.

\begin{figure}
\epsscale{1.11}
\centering
\plotone{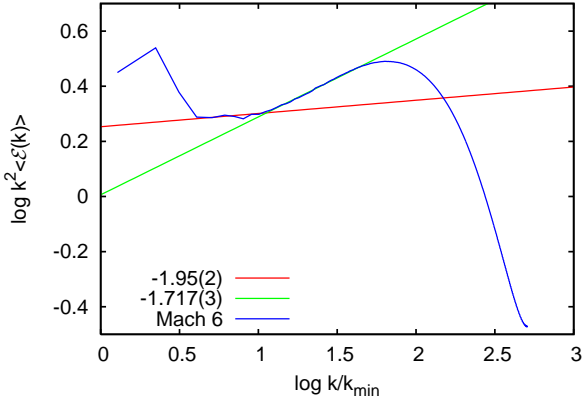}
\caption{Velocity power spectrum compensated by $k^2$. Note a large-scale excess of 
power at $\ell\in[256, 1024]\Delta$ due to external forcing, a short straight section 
in the uncontaminated inertial subrange $\ell\in[40, 256]\Delta$, and a small-scale 
excess at $\ell<40\Delta$ due to the bottleneck phenomenon. The straight lines represent 
the least-squares fits to the data for $\log k/k_{min}\in[0.6,1.1]$ and 
$\log k/k_{min}\in[1.2,1.7]$.}
\label{vpow}
\end{figure}
\begin{figure}
\epsscale{1.11}
\centering
\plotone{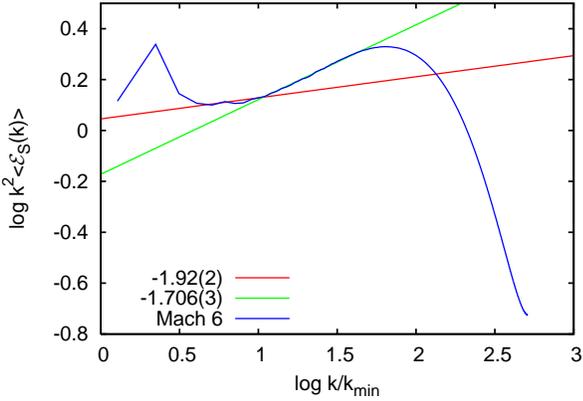}
\caption{Same as Fig. \ref{vpow}, but for the solenoidal velocity component.}
\label{vpows}
\end{figure}
\begin{figure}
\epsscale{1.11}
\centering
\plotone{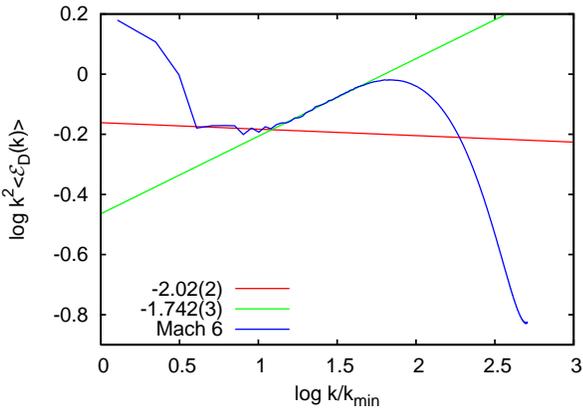}
\caption{Same as Fig. \ref{vpow}, but for the dilatational velocity component.}
\label{vpowc}
\vspace{3mm}
\end{figure}

Our reference velocity power spectrum from the $1024^3$ simulation is repeated in 
Fig.~\ref{vpow} with the best-fit linear approximations. It follows a power law with
an index $\beta=1.95\pm0.02$ within the range of scales $\ell\in[40,256]\Delta$ and 
has a shallower slope, $\beta_b=1.72$, for the bottleneck bump. After Helmholtz 
decomposition, ${\cal E}(k) = {\cal E}_S(k)+{\cal E}_D(k)$, the spectra for the 
solenoidal and dilatational components show the inertial range power indices of 
$\beta_S=1.92\pm0.02$ and $\beta_D=2.02\pm0.02$, respectively, see Figs.~\ref{vpows} 
and \ref{vpowc}. The difference in the slopes of ${\cal E}_S(k)$ and ${\cal E}_D(k)$ is 
about $5\;\sigma$, i.e. significant.
Both spectra display flattening due to the bottleneck with indices of $\beta_{b,S}=1.70$ and 
$\beta_{b,D}=1.74$ in the near-dissipative range. The fraction of energy in dilatational 
modes quickly drops from about $50$\% at $k=k_{min}$ down to $30$\% at 
$k/k_{min}\approx50$ and then returns back to a level of $45$\% at the Nyquist frequency. 

In summary, the inertial range scaling of the velocity power spectrum in highly compressible
turbulence tends to be closer to the Burgers scaling with a power index of $\beta=2$
rather than to the Kolmogorov $\beta=5/3$ scaling suggested for the mildly 
compressible transonic simulations by \citet{porter..02}. The inertial range scaling 
exponents of the power spectra for the solenoidal and dilatational velocities are not 
the same, with the latter demonstrating a steeper Burgers scaling, containing about 
$30$\% of the total specific kinetic energy, and being responsible for up to $35$\% 
of the global dissipation rate (see Section \ref{glb}).

\subsection{Velocity Structure Functions} \label{vsf}
\begin{figure}
\epsscale{1.125}
\centering
\plotone{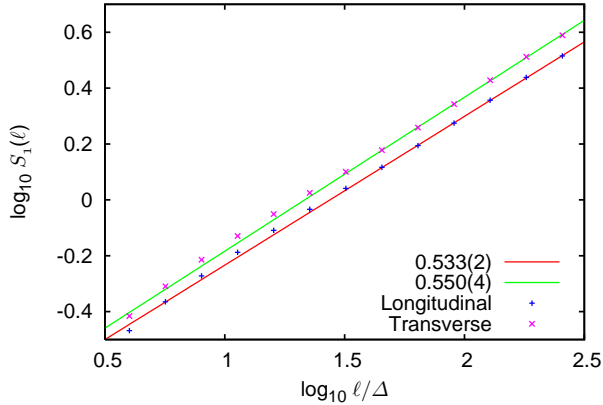}
\caption{First-order velocity structure functions and the best-fit power laws for 
$\log_{10}(\ell/\Delta)\in[1.5,2.5]$.}
\label{sf1}
\end{figure}

\begin{figure}
\epsscale{1.125}
\centering
\plotone{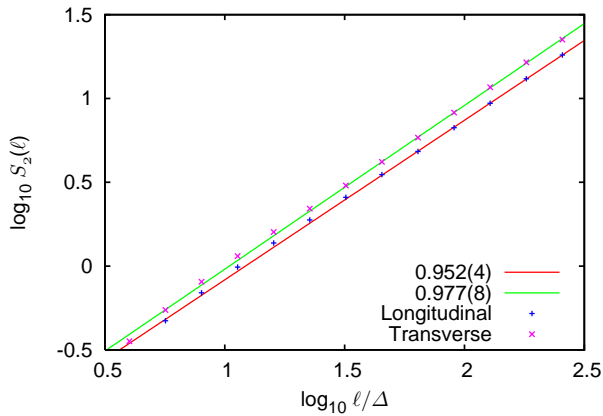}
\caption{Same as Fig. \ref{sf1}, but for the second-order structure functions.}
\label{sf2}
\end{figure}

\begin{figure}
\epsscale{1.125}
\centering
\plotone{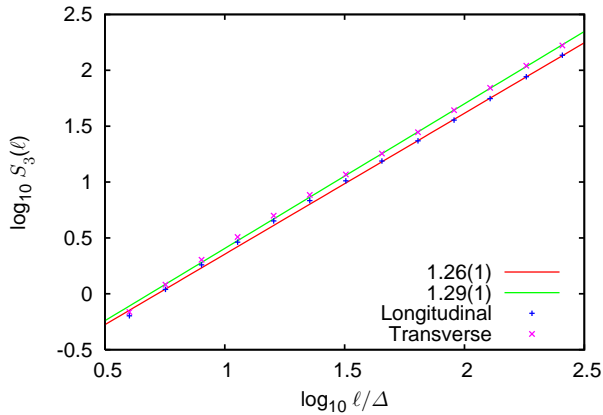}
\caption{Same as Fig. \ref{sf1}, but for the third-order structure functions.}
\label{sf3}
\end{figure}

To substantiate this result, we studied the scaling properties of the velocity structure 
functions \citep[e.g.,][]{monin.75}
\begin{equation}
S_p(\ell)\equiv\left<\left|\pmb{u}(\pmb{r}+\pmb{\ell})-\pmb{u}(\pmb{r})\right|^p\right> 
\label{sfdef}
\end{equation}
of orders $p=1$, $2$, and $3$, where the averaging $\left<\ldots\right>$ is taken over 
all positions $\pmb{r}$ and all orientations of $\pmb{\ell}$ within the computational domain. 
Both longitudinal ($\pmb{u}\parallel\pmb{\ell}$) and transverse ($\pmb{u}\perp\pmb{\ell}$) 
structure functions can be well approximated by power laws in the inertial range
\begin{equation}
S_p(\ell)\propto \ell^{\zeta_p},
\end{equation}
 see Figures~\ref{sf1}-\ref{sf3}.
The low-order structure 
functions are less susceptible to the bottleneck contamination and might be better
suited for deriving the scaling exponents \citep{dobler...03}.\footnote{Note, however, that
the bottleneck corrections grow with the order $p$ \citep{falkovich94} and influence the
structure functions in a nonlocal fashion, mixing small- and large-scale information
\citep{dobler...03,davidson.05}.} Let us now check how close the exponents derived from
the data are to the K41 prediction $\zeta_p=p/3$.

It is natural to begin with the second-order structure functions, since 
the velocity power spectrum ${\cal E}(k)$ is the Fourier transform of $S_2(\ell)$
and therefore $\beta=\zeta_2+1$. 
The best-fit second-order exponents measured for the range of scales $\ell$ between 
$32\Delta$ and $256\Delta$, $\zeta^{\parallel}_2=0.952\pm0.004$ and 
$\zeta^{\perp}_2=0.977\pm0.008$, are substantially larger than the K41
value of $2/3$ and, within the statistical uncertainty, agree with our measured 
velocity power index $\beta=1.95\pm0.02$.
In homogeneous isotropic incompressible turbulence, $S^{\perp}_2$ is uniquely determined 
by $S^{\parallel}_2$ (and vice versa) via the first \citet{karman.38} relation,
\begin{equation}
S^{\perp}_2=S^{\parallel}_2+\frac{\ell}{2}S^{\parallel\prime}_2. 
\end{equation}
For the K41 inertial range velocity scaling, this translates into 
\begin{equation}
S^{\perp}_2=\frac{4}{3}S^{\parallel}_2. 
\end{equation}
At Mach 6, the 2nd-order longitudinal and transverse structure functions 
are also approximately parallel to each other in the $(\log S,\; \log \ell)$-plane,
but the offset between them, $S^{\perp}_2/S^{\parallel}_2\approx 1.27$, is 
somewhat smaller than the K41 value of $4/3$.
This suggests that in highly compressible regimes some generalization of the first
K\'arm\'an-Howarth relation still remains valid, even though a substantial fraction
of the specific kinetic energy (up to $1/3$ at ${\cal M}=6$) is concentrated in 
the dilatational component \citep[cf.][]{moyal51,samtaney..01}.

\begin{figure}
\epsscale{1.125}
\centering
\plotone{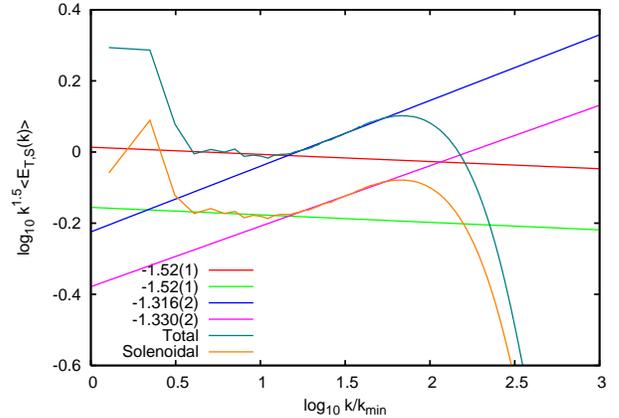}
\caption{Time-average total kinetic energy spectrum $E(k)$ and solenoidal 
kinetic energy spectrum $E_S(k)$, both compensated by $k^{3/2}$. The straight 
lines represent the least-squares fits to the data for $\log k/k_{min}\in[0.5,1.2]$ 
and $\log k/k_{min}\in[1.2,1.8]$. The inertial range slopes 
of the total and the solenoidal kinetic energy spectrum are the same within 
the statistical errors, which is not the case for the velocity power spectra. 
The dilatational component (not shown) also has the same slope.}
\label{eki}
\vspace{5mm}
\end{figure}

The first-order exponents $\zeta^{\parallel}_1=0.533\pm0.002$ and 
$\zeta^{\perp}_1=0.550\pm0.004$ are again significantly larger than the K41 index 
of $1/3$ but also considerably smaller than the index of $1$ for the Burgers model 
\citep[e.g.,][]{frisch.01}. Observationally determined exponents for the molecular 
cloud turbulence are normally found within the range from $0.4$ to $0.8$
\citep{brunt...03}, in reasonable agreement with our result.

We find that the third-order scaling exponents $\zeta^{\parallel}_3=1.26\pm0.01$ and 
$\zeta^{\perp}_3=1.29\pm0.01$ are also noticeably off from $\zeta_3=1$ predicted by K41
in the incompressible limit.
Our measurements for the three low-order exponents roughly agree with the estimates 
$\zeta^{\perp}_1\approx0.5$, $\zeta^{\perp}_2\approx0.9$, and $\zeta^{\perp}_3\approx1.3$
obtained by \citet{boldyrev..02} from numerical simulations of isothermal Mach 10 MHD 
turbulence at a resolution of $500^3$ points. A rigorous result $\zeta_3=1$ holds for 
incompressible Navier-Stokes turbulence \citep[][``$4/5$'' law]{kolmogorov41c} and for 
incompressible MHD \citep{politano.98}. Strictly speaking, $\zeta_3=1$ is proved only 
for the longitudinal structure function, $S^{\|}_3$, \citep{kolmogorov41c} and for certain 
mixed structure functions \citep{politano.98}, and in both cases the absolute value of 
the velocity difference is {\em not} taken (cf. eq. [\ref{sfdef}]), but still we know from
numerical simulations that $\zeta_3\approx1$ in a compressible transonic regime for the 
absolute velocity differences, as in equation (\ref{sfdef}) \citep{porter..02}.

The fact that in supersonic turbulence $\zeta_3\ne1$ should be accounted for in 
extensions of incompressible turbulence models into the strongly compressible
regime. One approach is to consider relative (to the third order) 
scaling exponents as more universal \citep{dubrulle94} and reformulate the models in terms 
of the relative exponents \citep[cf.][]{boldyrev02,boldyrev..02}. In many low-resolution
simulations, exploiting the so-called extended self-similarity hypothesis
\citep{benzi.....93,benzi....96}, the foundations of which are not yet fully understood 
\citep[see, e.g.,][]{sreenivasan.05}, is the only resort to at least measure the relative
exponents. Another approach is to consider statistics of mixed quantities, like 
$\rho^{\eta}\pmb{u}$, in place of pure velocity statistics naively inherited from 
incompressible regime studies. We shall discuss 
the options for the former approach below and then return to the latter in Sections
\ref{eps} and \ref{mdl}.

The {\em relative} scaling exponents $Z_p\equiv\zeta_p/\zeta_3$ of the first and the second 
order we obtain from our simulations are still somewhat higher than their incompressible 
nonintermittent analogues 1/3 and 2/3; $Z^{\parallel}_1 = 0.42$, $Z^{\perp}_1 = 0.43$, 
and $Z^{\parallel}_2 \approx Z^{\perp}_2 = 0.76$. This situation is similar to a small
discrepancy between the K41 predictions and experimental measurements that stimulated 
\citet{obukhov62} and \citet{kolmogorov62} to supplement the K41 theory by 
``intermittency corrections.'' If we follow a similar path in our strongly 
compressible case and estimate the ``corrected'' exponents using the \citet{boldyrev..02}
formula, which is similar to the \citet{she.94} model, but assumes that the most singular 
dissipative structures are not one-dimensional vortex filaments but rather two-dimensional 
shock fronts, 
\begin{equation}
Z_p=\frac{p}{9}+1-\left(\frac{1}{3}\right)^{p/3},
\label{b02}
\end{equation}
we get the exponents $Z_1=0.42$ and $Z_2=0.74$ that are very close to our estimates.
One can argue though that at high Mach numbers, the dilatational velocities constitute 
a substantial part of the total specific kinetic energy, and therefore, a double set of
singular velocity structures and their associated dynamics would lead to a compound Poisson 
statistic instead of a single Poisson process incorporated in equation (\ref{b02}) 
(see She \& Waymire 1995). To check whether this is indeed true or not, one needs to collect 
information about higher order statistics from simulations at resolution better than
$1024^3$.

\subsection{Kinetic Energy Spectrum}\label{eps}
To explore the second option outlined in Section 3.4, we considered the 
energy-spectrum function $E(k)$ built on $\pmb{w}\equiv\sqrt{\rho}\;\pmb{u}$
\citep[e.g.,][]{lele94} in addition to ${\cal E}(k)$ that depends solely on the 
velocity $\pmb{u}$ and is traditionally used for characterizing incompressible 
turbulence. We defined the kinetic energy spectral density $E(\pmb{k})$ as
the square of the Fourier coefficient
\begin{equation}
E(\pmb{k})\equiv\frac{1}{2}\left|\pmb{\tilde{w}}(\pmb{k})\right|^2
\end{equation}
determined by the Fourier transform
\begin{equation}
\pmb{\tilde{w}}(\pmb{k})=\frac{1}{(2\pi)^3}\int_{\cal V}\pmb{w}(\pmb{x})e^{-2\pi i 
\pmb{k}\cdot\pmb{x}}d\pmb{x}.
\end{equation}
Then we used the standard definition of the three-dimensional kinetic 
energy spectrum function
\begin{equation}
E(k)\equiv\int_{\tilde{\cal V}} E(\pmb{k})\delta(|\pmb{k}|-k)d\pmb{k},
\end{equation}
where $\delta(k)$ as before is the $\delta$-function. To avoid a possible 
terminological confusion, in the following we always refer to 
${\cal E}(k)$ as the velocity power spectrum and to $E(k)$ as the kinetic 
energy spectrum
\begin{equation}
E\equiv\int E(k) d k = \frac{1}{2}\left<\rho\pmb{u}^2\right>.
\end{equation}
\begin{figure}
\epsscale{1.15}
\centering
\plotone{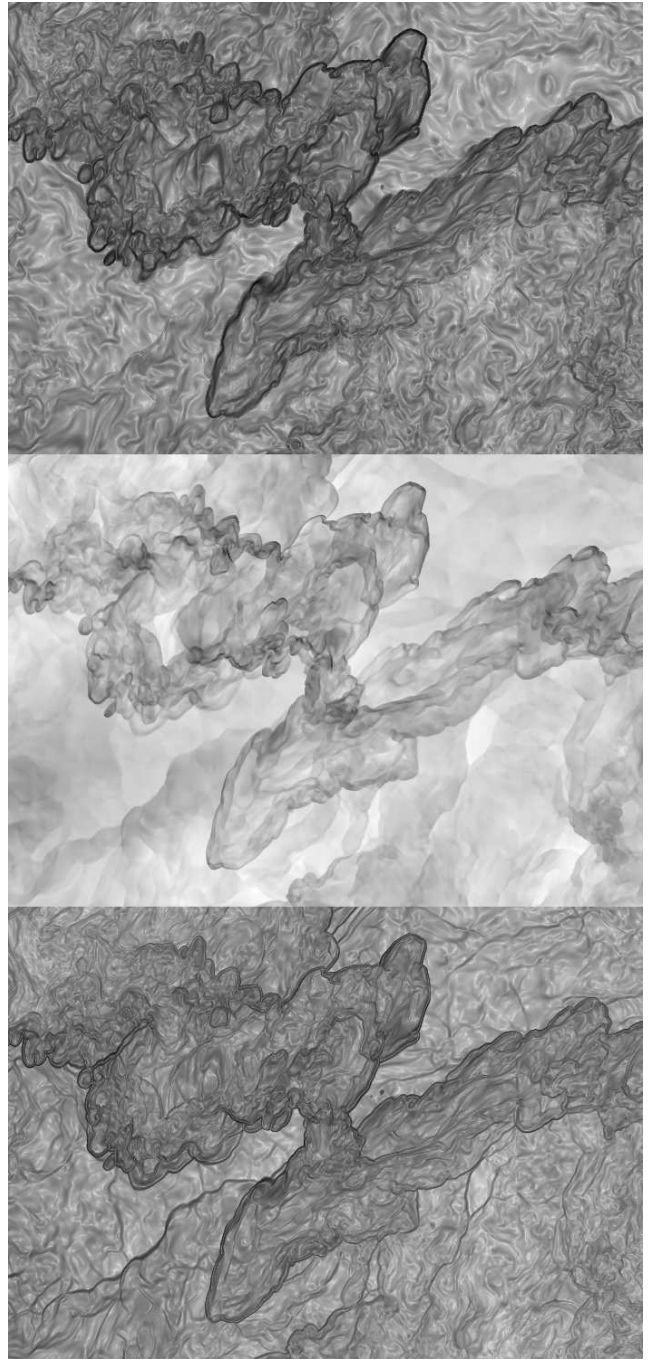}
\caption{Enstrophy ($|\pmb{\nabla}\times\pmb{u}|^2$; {\em top}), 
density ({\em middle}), and ``denstrophy'' 
($|\pmb{\nabla}\times(\sqrt{\rho}\pmb{u})|^2/\rho$; {\em bottom}) 
distributions in a slice through the center of the subvolume of the 
computational domain at $t=7t_d$. The logarithmic gray-scale ramp 
is used to show the highest values in black and the lowest values 
in white.}
\label{sli}
\end{figure}

\begin{figure}
\epsscale{1.15}
\centering
\plotone{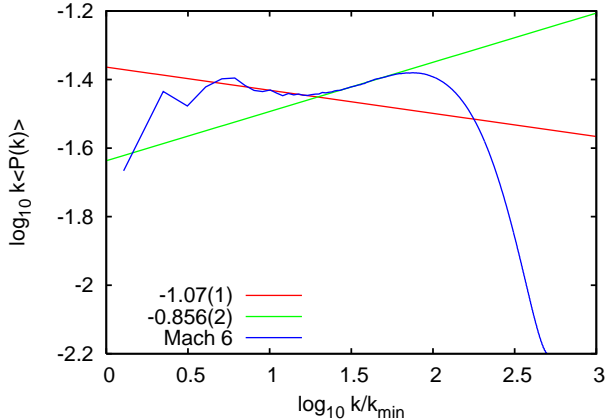}
\caption{Time-average density power spectrum compensated by $k$. The straight lines represent 
the least-squares fits to the data for $\log k/k_{min}\in[0.6,1.3]$ 
and $\log k/k_{min}\in[1.4,1.7]$.}
\label{dpow}
\end{figure}

At low Mach numbers, the scaling exponents of $E(k)$ and ${\cal E}(k)$ are nearly the
same. However, while ${\cal E}(k)$ does not change much as the Mach number approaches unity
\citep[e.g.,][]{porter..02}, the slope of $E(k)$ gets shallower already 
by ${\cal M}=0.9$ \citep[][based on simulations with $64^3$ collocation points]{kida.92}. 
Since at low Mach numbers
the spectra are dominated by the solenoidal components, and the solenoidal velocity power
spectrum hardly depends on the Mach number, the departure of the slope of $E(k)$ from the slope
of ${\cal E}(k)$ is mostly due to the sensitivity of the solenoidal part of the kinetic
energy spectrum to the density--vorticity correlation that is felt already around 
${\cal M}\sim0.5$. Shallow slopes for $E(k)$ with power indices around $-3/2$
were also detected in MHD simulations of highly supersonic super-Alfv\'enic
turbulence \citep[][$512^3$ grid points]{li...04}.

In the inertial range, the time-averaged kinetic energy spectrum scales as 
$E(k)\sim k^{-\beta}$ with $\beta=1.52\pm0.01$ (see Fig.~\ref{eki}). 
The spectrum displays flattening at high wavenumbers due to the bottleneck 
effect, which is similar to what we have already seen in the velocity power 
spectra ${\cal E}(k)$. The flat part scales roughly as $\sim k^{-1.3}$ and 
occupies the same range of wavenumbers as in the velocity spectrum. Since 
the kinetic energy spectrum is sensitive to the rather sporadic activity of 
density fluctuations, time-averaging is essential to get a robust estimate 
for the power index.

We performed a decomposition of $\pmb{w}\equiv\sqrt{\rho}\pmb{u}$ into the solenoidal and 
dilatational parts $\pmb{w}_{S,D}$ such that 
$\pmb{\nabla}\cdot\pmb{w}_S=\pmb{\nabla}\times\pmb{w}_D\equiv 0$, and computed the energy
spectra for both components, $E(k)=E_S(k)+E_D(k)$. The inertial range spectral exponents 
for the solenoidal and dilatational components are the same, $\beta=1.52\pm0.01$ for
both, and also coincide with the slope of $E(k)$. This remarkable property of $E(k)$ 
suggests that in the supersonic regime\footnote{In our Mach 6 simulations at $1024^3$, the
sonic scale $\ell_s$ such that $u(\ell_s)\sim c=1$ is located in the middle of the 
bottleneck bump, and thus, the velocities within the inertial range are supersonic.
One could possibly detect a break in the velocity power spectrum from a steep Burgers-like
supersonic scaling at low $k$ to a shallow Kolmogorov-like transonic scaling at higher 
wavenumbers in Mach~6 simulations with resolution of $4000^3$ grid 
points or higher. Note, that as we show later in Section~\ref{mdl}, the power spectrum 
of $\rho^{1/3}\pmb{u}$ would not show 
such a break and would instead approximately follow the Kolmogorov 5/3 law all over 
the inertial interval.} we are dealing with a single compressible cascade of 
kinetic energy where the density fluctuations provide a tight coupling between the 
solenoidal and dilatational modes of $\pmb{w}$. This picture is similar in spirit to 
that discussed by \citet[][in their Section 3.7]{kornreich.00}. However, in contrast to 
\citet{kornreich.00}, our simulations demonstrate that nonlinear HD instabilities 
are heavily involved in the kinetic energy transfer through the hierarchy of scales. 
The association of $\pmb{w}$ with the kinetic 
energy distribution as a function of scale makes this quantity a better candidate than 
the pure velocity $\pmb{u}$ for employing in compressible cascade models, since $\pmb{w}$ 
uniquely represents the physics of the cascade and there is no need to
track the variation of ${\cal E}_D(k)/{\cal E}(k)$ as a function of wavenumber in the
inertial subrange in addition to variations with the rms Mach number.

The total kinetic energy power is dominated by the solenoidal component $\pmb{w}_S$ over 
the whole spectrum.
Within the inertial range, $E_S(k)$ contributes about $68$\% of the total power, then $\sim66$\%
in the bottleneck-contaminated interval, and up to $\sim74$\% further down at the Nyquist 
frequency. Compare with the solenoidal part of the {\em velocity} power, ${\cal E}_S(k)$,
that constitutes about $65$\%$-70$\% within the inertial range and only $55$\% of the total power 
at the Nyquist frequency (see Section \ref{vps}). 
\begin{figure*}
\centering
\epsscale{1.15}
\plotone{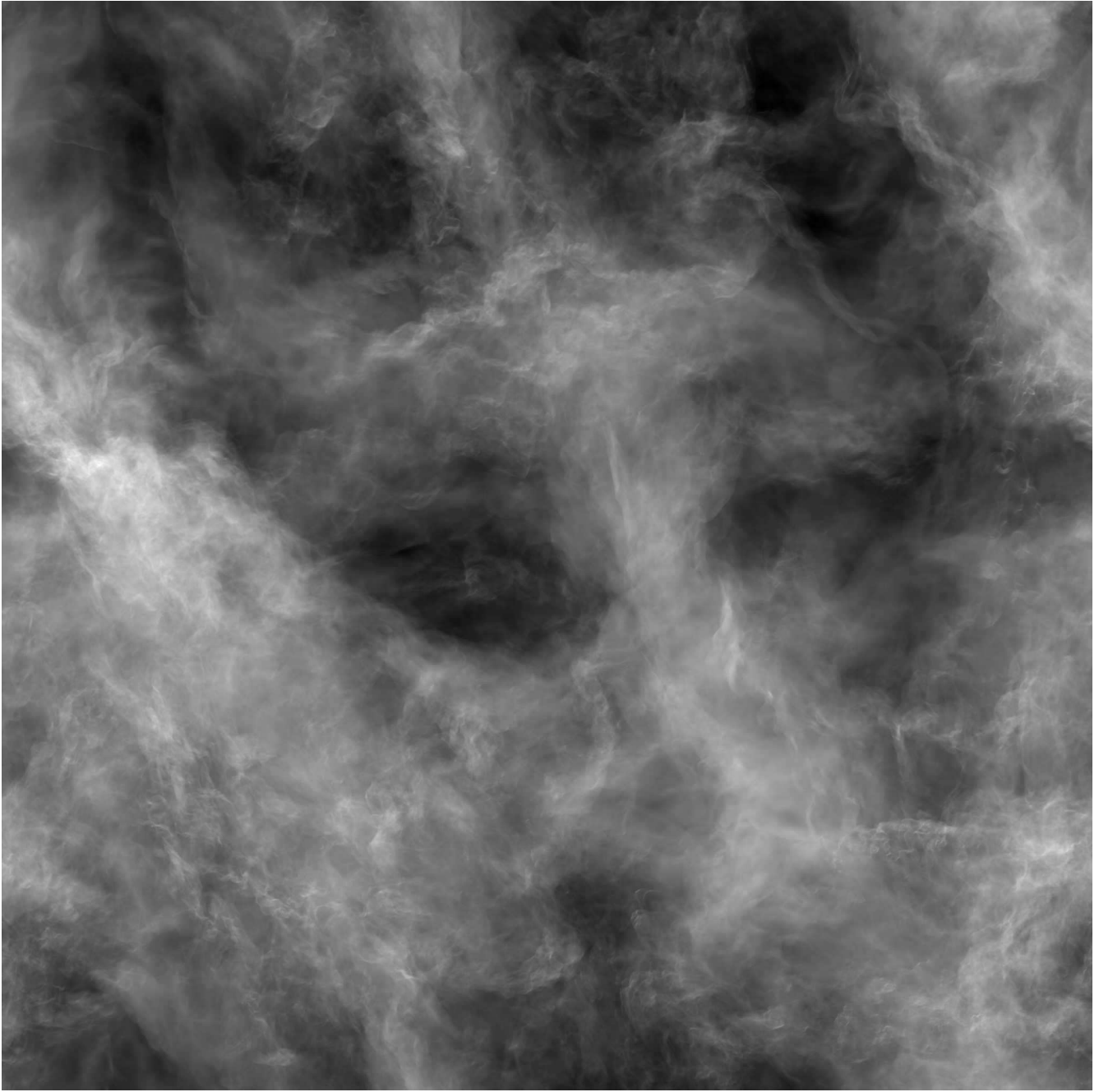}
\caption{Logarithm of the projected density from a snapshot of the AMR simulation
with effective grid resolution of $2048^3$ zones. The standard linear gray-scale 
ramp shows the highest density peaks in white and the most underdense voids in black.}
\label{fullp}
\end{figure*}

\begin{figure*}
\epsscale{1.15}
\centering
\plotone{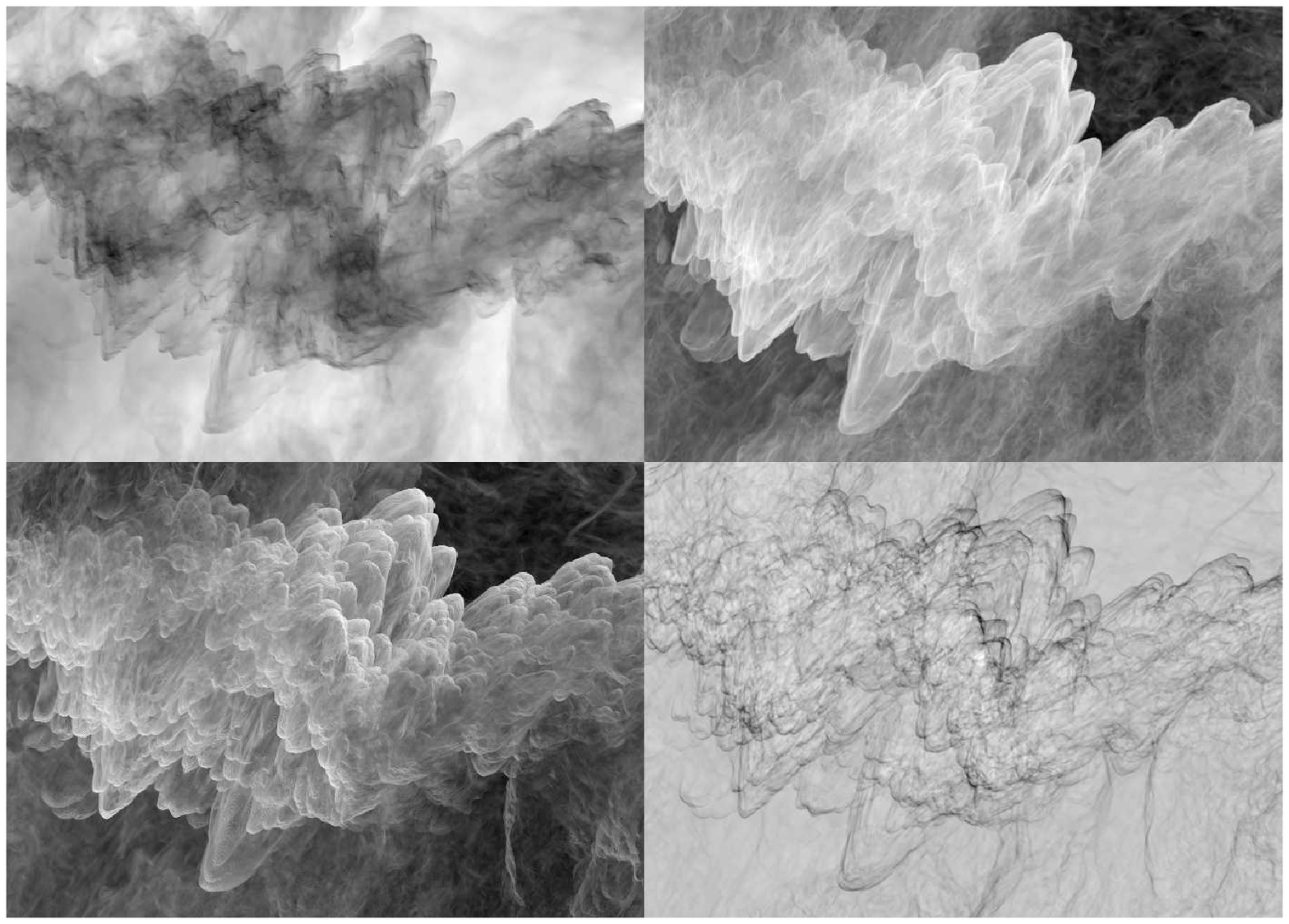}
\caption{Coherent structures in Mach 6 turbulence at resolution of $1024^3$. 
Projections along the minor axis of a subvolume of $700\times500\times250$ 
zones for the density ({\em top left}),
the enstrophy ($|\pmb{\nabla}\times \pmb{u}|^2$; {\em top right}), 
the dissipation rate (the integrand $|\pmb{\Phi\cdot u}|$ in eq.~\ref{13}; 
{\em bottom left}), and the dilatation ($\pmb{\nabla}\cdot \pmb{u}$; 
{\em bottom right}). The logarithmic gray-scale ramp shows the lower values 
as dark in all cases except for the density. The inertial subrange structures 
correspond to scales between $40$ and $250$ zones and represent a fractal 
with $D_m\approx2.4$. The most singular structures in the dissipation range 
($\ell < 30\Delta$) are shocks with fractal dimension $D_m=2$.}
\label{morph}
\end{figure*}

Figure \ref{sli} illustrates the difference
in structures seen in the enstrophy field ({\em top}) and in the ``denstrophy'' 
field ($|\pmb{\nabla}\times(\sqrt{\rho}\pmb{u})|^2/\rho$, {\em bottom}) that helps 
to understand the small-scale excess of power in $E_S(k)$ with respect to ${\cal E}_S(k)$. 
While the 
corrugated shock surfaces (U-shapes), which are also the regions of very strong shear, are 
seen as dark wormlike structures of excess enstrophy and denstrophy in both the top and 
the bottom panels of Fig. \ref{sli}, nearly planar shock fronts that carry a negligible 
amount of shear (this is also why they remain planar) are clearly missing in the enstrophy 
plot. Since the denstrophy field contains a greater number of sharper small-scale structures, 
it should be expected that $E_S(k)$ carries more small-scale power than ${\cal E}_S(k)$.
Overall, the structures captured as intense by the denstrophy field closely follow
the regions of high energy dissipation rate (see the integrand in eq. [\ref{13}]
and Fig.~\ref{morph}, {\em bottom left}).

\subsection{Density Power Spectrum}\label{dps}
The power spectrum of the gas density shows
a short straight section with a slope of $-1.07\pm0.01$ in the range of scales from 
$250\Delta$ down to $40\Delta$ followed by flattening due to a power pileup 
at higher wavenumbers, see Fig.~\ref{dpow}.
Similar power-law sections, and excess of power in the same wavenumber ranges
were seen earlier in the velocity and the kinetic energy power spectra, 
Figs.~\ref{vpow} and \ref{eki}. 

At the resolution of $512^3$, the bottleneck bump at high wavenumbers and the
external forcing at low $k$ leave essentially no room for the uncontaminated 
inertial range in $k$-space, even though the density spectrum at $512^3$ 
(not shown) also has a straight power-law section. At $512^3$, the slope of 
the density power spectrum, $-0.90\pm0.02$, is substantially shallower than 
$-1.07$ at $1024^3$. Thus, spectral index estimates based on low-resolution 
simulations bear large uncertainties due to the bottleneck contamination. 
We also note that the time-averaging over many snapshots is essential to get 
the correct slope for the density power spectrum, since the density exhibits 
strong variations on very short (compared to $t_d$) time-scales. 
The spectrum tends to get shallower after collisions of strong shocks, when the 
PDF's high-density wing rises above its average lognormal representation.

There are good reasons to believe that in weakly compressible isothermal flows 
the three-dimensional density power spectrum scales in the inertial subrange as 
$\sim k^{-7/3}$ \citep{bayly..92}. 
Our $512^3$ transonic simulation at ${\cal M}=1$ with purely 
solenoidal driving (not shown) returned a time-average power index of 
$-1.6\pm0.1$, which is probably still too shallow due to the unresolved inertial 
range (see also Section \ref{vps}). An index of $-1.7$ at ${\cal M}=1.2$ 
was obtained by \citet{kim.05} at the same resolution (but with a more diffusive 
solver). Our simulations at higher resolution for ${\cal M}= 6$ give a slope 
of approximately $-1.1$. At even higher Mach numbers, the spectra tend to 
flatten further. From the continuity arguments, there should exist a Mach number 
value ${\cal M}_{41}$ such that the power index of the density spectrum is 
exactly $-5/3$, i.e. coincides with the Kolmogorov velocity spectrum power index. 
Since apparently ${\cal M}_{41}\lesssim1$, subsonic (but not too subsonic) compressible 
turbulent flows should have near-Kolmogorov scaling, even for the density power 
spectrum \citep[cf.][]{armstrong..95}.

\subsection{Coherent Structures in Supersonic Turbulence}
Let us now look more carefully at the morphology of those structures in supersonic
turbulence that determine its scaling properties. The very first look at the projected
density field shown in Figure~\ref{fullp} reveals a plethora of filamentary structures 
very similar to what one usually sees as cirrus clouds in the sky. This is roughly what 
one can directly observe, albeit with a somewhat poorer resolution, in the intensity maps 
of nearby molecular clouds and in the WC-type turbulent stellar winds from the
Wolf-Rayet stars.

If one zooms in on a subvolume and takes a slab that is 5 times thinner than the whole
domain, one will start seeing the elements of that cloudy structure, namely, numerous 
nested U- and V-shaped  bow shocks or ``Mach cones'' \citep{kritsuk..06}. 
Figure~\ref{morph} shows an example of coherent structures that form in supersonic
flows as a result of development and saturation of linear and nonlinear instabilities 
in shear layers formed by shock interactions.\footnote{The (incomplete) list of potential 
candidates includes the Kelvin-Helmholtz instability \citep{helmholtz1868},
the nonlinear supersonic vortex sheet instability \citep{miles57}, and the
nonlinear thin shell instability \citep{vishniac94}.} The instabilities assist in
breaking the dense post-shock gas layers into supersonically moving fragments. 
Due to their extremely high density contrast with the surrounding medium, the
fragments behave as effectively solid obstacles with respect to the medium.
These coherent structures 
are transient and form in supersonic collisions of counter-propagating flow patches or 
``blobs'' \citep{kritsuk.04,heitsch....05,kritsuk..06}. Then they dissipate and similar
structures appear again and again at other locations. The nested V-shapes can be best 
seen in slabs with a finite thickness of the order of their characteristic size (i.e. 
$\sim200\Delta$ in our $1024^3$ run) that belongs to the resolved inertial subrange of 
scales. They are less noticeable in full-box projections where multiple structures of 
the same sort tend to overlap along the line of sight as in Figure~\ref{fullp}. 
Since the ``cones'' are essentially three-dimensional, they can also be hardly 
seen in thin slices, see Fig. \ref{sli}. Patterns of the same morphology were
identified in the peripheral parts of the M1-67 nebula observed in H$_{\alpha}$ 
emission by the {\em Hubble Space Telescope} \citep{grosdidier....01}.

\subsection{Fractal Dimension of the Mass Distribution\label{fr}}
It is known from observations and experiments that the distribution of dissipation 
in turbulent flows is intrinsically intermittent and this intermittency bears a 
hierarchical nature. This fact is in apparent contradiction with the K41 theory 
that assumes that the rate of dissipation is uniform in space and constant in time. 
\citet{mandelbrot74} introduced a new concept of the ``intrinsic fractional dimension 
of the carrier,'' $D$, that characterizes the geometric properties of a subset of the
whole volume of turbulent flow where the bulk of intermittent dissipation occurs.
He also studied the relation between $D$ and scaling properties of turbulence.
In particular, he suggested that isosurfaces of scalars (such as concentration or 
temperature) in turbulent flows with Burgers and Kolmogorov statistics are 
best described by fractal dimensions of $3-1/2$ and $3-1/3$, respectively 
\citep{mandelbrot75}. In the following we focus on similar questions with
respect to supersonic turbulence.

In turbulence research there are several approaches to assess the flow geometry
quantitatively.
Indirectly, the information on fractional dimension of dissipative structures
in turbulent flows can be obtained as a by-product of application of phenomenological 
cascade models, e.g. the hierarchical structure (HS) model developed by 
\citet{she.94} and \citet{dubrulle94} for incompressible Navier-Stokes 
turbulence and later extended to supersonic flows by \citet{boldyrev02}.
The HS model is based on the assumption of log-Poisson statistics for the energy
dissipation rate and provides an estimate for fractal dimension, $D$, of the most 
singular structures based on the detailed knowledge of high-order {\em velocity} 
structure functions \citep[e.g.,][]{kritsuk.04,padoan...04}.
While the HS model appears to reproduce experimental data quite well in a rather
diverse set of applications extending far beyond the limits of the original study 
of fully developed turbulence \citep[e.g.,][]{liu....04}, the uncertainty in the scaling
properties of high-order velocity statistics remains large. To nail down the $D$ value
thus obtained from numerical experiments, one has to go beyond the current 
resolution limits to resolve the inertial subrange dynamics sufficiently well 
and to provide a statistical data sample of a reasonably large size. That would 
reduce the effects of the statistical noise and anisotropies which tend to strongly 
contaminate the high-order velocity statistics \citep[e.g.,][]{porter..02}. We leave a more 
detailed discussion of the HS model to a follow-up paper and focus here instead on 
a direct measurement of the fractal dimension of the density field using conventional 
techniques. Note, however, that there is no trivial physical connection between the 
fractal dimensions measured by these two techniques, i.e. they do not necessarily refer 
to the same fractal object since one is based exclusively on the velocity information,
while the other is based on the density information.

For physical systems in three-dimensional space, like supersonic turbulent boxes,
it is often easier to directly estimate the mass dimension that contains information 
about the density scaling with size. To do so, we select a sequence of box sizes 
$\ell_1 > \ell_2 > ... > \ell_n$ and cover high-density peaks with concentric
cubes of those linear sizes. Denoting $M(\ell_i)$ as the mass contained inside
the box of size $\ell_i$, one can plot $\log M(\ell_i)$ versus $\log \ell_i$ to
reveal a scaling range where the points follow a straight line,
\begin{equation} 
M(\ell)\propto\ell^{D_m}.
\end{equation} 
The slope of the line, $D_m$, then gives an estimate of the mass dimension for 
the density distribution. Note that the centers of the box hierarchies cannot 
be chosen arbitrarily. They must belong to the fractal set, or else, in the 
limit $\ell\to 0$, one would end up with boxes containing essentially no mass. 
In practice, we selected as centers the positions where the gas density 
is higher or equal to one-half of the density maximum for a given snapshot. 
We then determined $D_m$ by taking an average over all centers from five 
statistically independent density snapshots separated from each other by roughly 
one dynamical time.

\begin{figure}
\epsscale{1.15}
\centering
\plotone{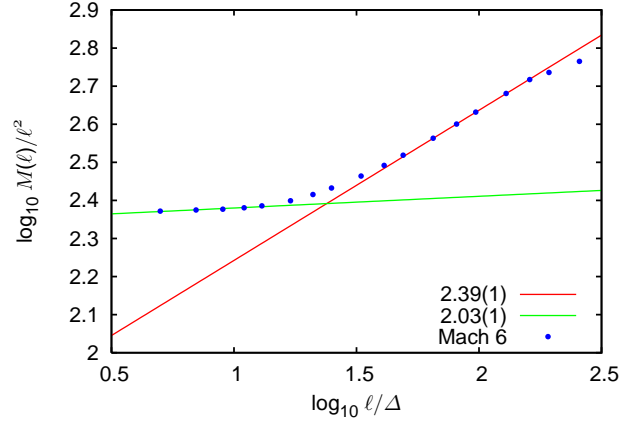}
\caption{Gas mass within a box of size $\ell$ normalized by $\ell^2$ as a function 
of the box size. The mass dimension $D_m$ of the density distribution is determined 
by the slope of this relationship. A horizontal line would correspond to $D_m=2$.
The data points represent averages over five density snapshots.
The straight lines are the least-squares fits to the data for $\log\ell/\Delta\in[0.5,1.2]$
and $\log\ell/\Delta\in[1.7,2]$.}
\label{fractal}
\vspace{3mm}
\end{figure}

Figure~\ref{fractal} shows the mass scaling with the box size $\ell$ 
compensated by $\ell^{-2}$ based on the procedure described above. 
The slope of the mass--scale dependence breaks from roughly 2 in the 
dissipation-dominated range of small scales to $\sim2.4$ in the inertial 
subrange that is free from the bottleneck effect. The transition point at 
$\ell\sim30\Delta$ coincides with the break point present in all power spectra
discussed above. On small scales, the geometry of the density field in 
supersonic turbulence is dominated by two-dimensional shock surfaces.
This is where the numerical dissipation dominates in PPM.

On larger scales, where the dissipation is practically negligible, 
the clustered shocks form a sophisticated pattern with fractal dimension 
$D_m=2.39\pm0.01$. Since the dynamic range of our simulations is
limited, we can only reproduce this scaling in a quite narrow interval from 
$\sim40\Delta$ to $\sim160\Delta$, and thus, the actual uncertainty of our 
estimate for $D_m$ in the inertial range can still be on the order of $\pm0.1$
or even larger. 
The break present in the ($\log M - \log \ell$) relation may indicate a 
change in intermittency properties of turbulence at the meeting point of the 
inertial and dissipation ranges \citep[see, e.g.][]{frisch95}.

\subsection{A Simple Compressible Cascade Model} \label{mdl}
More than half a century ago, \citet{vonweizs51} introduced a phenomenological model 
for three-dimensional compressible turbulence with an intermittent, scale-invariant 
hierarchy of density fluctuations described by a simple equation that relates the 
mass density at two successive levels to the corresponding scales through a universal 
measure of the degree of compression, $\alpha$,
\begin{equation}
\frac{\rho_{\nu}}{\rho_{\nu-1}}=\left(\frac{\ell_{\nu}}{\ell_{\nu-1}}\right)^{-3\alpha}.
\label{vws}
\end{equation}
The only free parameter of the model is the geometrical factor $\alpha$ which takes the
value of $1$ in a special case of isotropic compression in three dimensions, $1/3$ for 
a perfect one-dimensional compression, and zero in the incompressible limit.

The kinetic energy supplied to the system at large scales is
being transferred through the hierarchy by nonlinear interactions.
\citet{lighthill55} 
pointed out that, in a compressible fluid, the mean {\em volume} energy
transfer rate $\rho u^2 u/\ell$ is constant in a statistical steady state, so that
\begin{equation}
u\sim (\ell/\rho)^{1/3}.
\label{ekc}
\end{equation}
From these two equations, assuming mass conservation, \citet{fleck96} derived a set of scaling 
relations for the velocity, specific kinetic energy, density, and mass:\footnote{Similar
ideas were discussed earlier by \citet{biglari.88,biglari.89} in a slightly different 
context of self-gravitating compressible fluids.}
\begin{equation}
u\sim \ell^{1/3+\alpha},
\label{s1}
\end{equation}
\begin{equation}
{\cal E}(k)\sim k^{-\beta}\sim k^{\,-5/3-2\alpha},
\end{equation}
\begin{equation}
\rho\sim \ell^{\;-3\alpha},
\label{de}
\end{equation}
\begin{equation}
M(\ell)\sim \ell^{D_m}\sim \ell^{\;3-3\alpha},
\label{dm1}
\end{equation}
where all the exponents depend on the compression measure $\alpha$. The compression
measure, in turn, is a function of the rms Mach number of the turbulent flow.
In the incompressible limit, ${\cal M}\to 0$ and $\alpha\to 0$. There are also scaling
relations for $v\equiv\rho^{1/3}u$ that follow directly from equation (\ref{ekc}) and extend the 
incompressible K41 velocity scaling into the compressible regime,\footnote{Note that
\citet{kolmogorov41a,kolmogorov41b} and \citet{obukhov41} only considered $p\leq 3$
and never went as far as to consider $p>3$ \citep[cf.][]{frisch95}.}
\begin{equation}
v^p=(\rho^{1/3}u)^p\sim \ell^{\;p/3}.
\label{mxd}
\end{equation}
These hint at a unique generalization of the velocity structure functions for
compressible flows,
\begin{equation}
{\cal S}_p(\ell)\equiv\left<\left|\pmb{v}(\pmb{r}+\pmb{\ell}) - 
\pmb{v}(\pmb{r})\right|^p\right>\sim \ell^{\;p/3},
\label{sf41}
\end{equation}
with ${\cal S}_3\sim\ell$ (see also the discussion in Section~\ref{vsf}).
The scaling laws expressed by equation (\ref{sf41}) should not necessarily be exact 
and, as the incompressible K41 scaling, may require ``intermittency corrections'' 
that will be addressed in detail elsewhere. Our discussion here is limited to low-order 
statistics ($p\leq3$) for which the ``corrections'' are supposedly small. 
The compression measure $\alpha$ could also depend on the order $p$ in 
addition to its Mach number dependency. However, as we show below, a linear 
approximation $\alpha=const$ built into this simple cascade model may prove reasonable 
for the low-order density and velocity statistics.

Before turning to properties of the density-weighted velocities $\pmb{v}$, let us
first assess the predictive power of Fleck's model using the statistics we
already derived. Since from the simulations we know the scaling of the first-order 
velocity structure functions, we can get an estimate of the geometric factor 
$\alpha$ for the Mach 6 flow using equation (\ref{s1}). 
Assuming the inertial range scaling, $S_1\sim\ell^{0.54}$, we get $\alpha\approx0.21$ 
and $D_m\approx2.38$ from equation (\ref{dm1}). This is consistent with our direct 
measurement of the mass dimension, $D_m\approx2.4$, for the same range of scales.

We can also do a consistency check based on the second-order statistics.
From the inertial range scaling of the second-order structure functions, we get 
${\cal E}(k)\sim k^{-1.97}$ and therefore $\alpha=0.15$. This value corresponds to the 
fractal dimension $D_m\approx2.55$ that is higher, but still reasonably close to our 
direct estimate. The discrepancy can in part be attributed to a deficiency of Fleck's
model that asymptotically approaches the space-filling K41 cascade in the limit of weak 
compressibility ($\alpha\to0$) and thus does not take the full account of the 
intermittency of the velocity field.

It is interesting to compare the interface dimension $D_i$ introduced by 
\citet{meneveau.90} for intermittent incompressible turbulence based on the 
so-called Reynolds number similarity
\begin{equation}
D_i=2+\zeta_1,
\end{equation}
where $\zeta_1$ as before is the exponent of the first-order velocity structure 
function, with the mass dimension $D_m=3-3\alpha$. Since both quantities 
characterize the same intermittent structures, they may well refer to the same 
fractal object in our compressible case. If one assumes $D_i=D_m=D$, keeping 
in mind that $\zeta_1=1/3+\alpha$ as follows from (\ref{s1}), it is easy to 
compute the compression parameter $\alpha=1/6\approx0.16(6)$ as well as the 
fractal dimension, $D=2.5$, that appear to be in accord with the original 
proposal by \citet{fleck96} and with the numbers listed above.

Finally, we can use the scaling exponents of the velocity power spectrum 
${\cal E}(k)\sim k^{-1.97}$ and of the kinetic energy spectrum $E(k)\sim k^{-1.52}$
to find $\alpha$ from the density scaling in equation (\ref{de}). We get 
\begin{equation}
\rho\sim\frac{E(k)}{{\cal E}(k)}\sim k^{0.45} \sim \ell^{\;-3\alpha},
\label{dm}
\end{equation}
and thus, $\alpha=0.15$, consistent with the previous estimate based on the second-order 
statistics.

Overall, within the uncertainties, Fleck's model appears to successfully reproduce 
the low-order velocity and density statistics from our numerical simulations
and supports our direct measurement of the mass dimension $D_m$ as well.
\begin{figure}
\epsscale{1.15}
\centering
\plotone{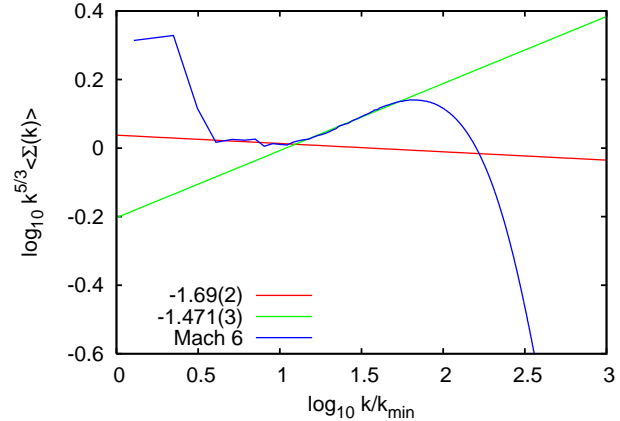}
\caption{Time-averaged power spectrum of the density-weighted velocity 
$\pmb{v}\equiv\rho^{1/3}\pmb{u}$ compensated by $k^{5/3}$. 
The straight lines represent the least-squares fits to the data 
for $\log k/k_{min}\in[0.5, 1.1]$ and $\log k/k_{min}\in[1.2, 1.8]$. 
The inertial subrange slope is in excellent agreement with the model 
prediction.}
\label{e41pow}
\vspace{3mm}
\end{figure}

Let us now check how well equation (\ref{mxd}) is satisfied in our simulations by
computing the power spectrum $\Sigma(k)$ of the density-weighted velocity 
$\pmb{v}$ as a $p=2$ diagnostic. We define $\Sigma(k)$ in exactly the same 
way as ${\cal E}(k)$ was defined in Section \ref{vps}, but substitute
$\pmb{v}$ in place of the velocity $\pmb{u}$. As can be seen from 
Figure~\ref{e41pow}, the inertial range power index, $\beta=1.69\pm0.02$, is
very close to the Kolmogorov value of $5/3$. 

To further check our conjecture expressed by equation \ref{sf41}, we measured the
first three scaling exponents for the modified structure functions ${\cal S}_p$
using a subsample of 35 snapshots evenly distributed in time through 
$t\in[6, 10]t_d$ using a sample of $2\times10^9$ point pairs per PDF per snapshot.
The resulting third-order exponents are very close to unity as expected,
$\zeta_3^{\perp}=1.01\pm0.02$ and $\zeta_3^{\|}=0.95\pm0.02$. This strongly suggests
that a relationship similar to Kolmogorov's ``4/5'' law may hold for compressible
flows as well.

Both the first-order exponents $\zeta_1^{\perp}=0.467\pm0.004$ and $\zeta_1^{\|}= 0.451\pm0.004$ 
and the second-order exponents $\zeta_2^{\perp}=0.80\pm0.01$ and 
$\zeta_2^{\|}= 0.76\pm0.01$ slightly deviate from the K41 $p/3$ scaling as would be
the case in intermittent turbulence. The corresponding relative 
(to the 3rd order) exponents $Z_1^{\perp}=0.46$ and $Z_1^{\|}=0.47$ and 
$Z_2^{\perp}=0.79$ and $Z_2^{\|}=0.80$ are also somewhat higher than their counterparts
for the velocity structure functions discussed in Section~\ref{vsf}, indicating
certain structural differences between the $\pmb{u}$ and $\pmb{v}$ fields.

Note, 
as the Mach number changes from ${\cal M}<1$ to ${\cal M}=6$, the slope of
the density power spectrum gets shallower from $-7/3$ to $-1.1$, but the
slope of the velocity power spectrum gets steeper from $-5/3$ to $-1.9$.
At the same time, the power spectrum of the mixed variable $\pmb{v}$ remains
approximately invariant, as predicted by the model. 

Our numerical experiments thus confirm one of the basic assumptions adopted 
in the compressible cascade model, namely, that equation (\ref{ekc}) for 
the energy transfer rate holds true. The first assumption concerning the properties of the
self-similar hierarchical density structure, equation (\ref{vws}) put forward by 
\citet{vonweizs51}, seems to be also satisfied in the simulations quite well,
at least to the first order.

\section{Discussion} \label{dscs}
The major deficiency of the numerical experiments discussed above is still their limited
spatial resolution that bounds the integral scale Reynolds numbers to values much smaller
than those estimated for the real molecular clouds. This hurdle apparently cannot
be overcome in the near future, but still the progress achieved in the past 15 years
in this direction is very impressive. 

The second important deficiency of our model is the lack of magnetic effects which are
known to be essential for star formation applications. This subject still remains a
topic for the future work awaiting the development of a high-quality MHD solver suitable
to modeling of supersonic flows at moderately high Reynolds numbers with
computational resources available today.

Another set of potential issues relates to the external driving force that is 
supposed to simulate the energy input by HD instabilities in real 
molecular clouds. The large-scale driving force we used in these simulations 
is not perfectly isotropic due to the uneven distribution of power between the 
solenoidal and dilatational modes (perhaps, a typical situation for the 
interstellar conditions). We also use a {\em static} driving force that could 
potentially cause some anomalies on timescales of many dynamical times. However, 
while strong anisotropies can significantly affect the scaling of high-order moments 
\citep{porter..02,mininni..06}, the departures from Kolmogorov-like scaling we 
observe in the lower order statistics appear to be too strong to be explained 
solely as a result of the specific properties of the driving. The sensitivity 
of our result to turbulence forcing remains to be verified with future 
high-resolution simulations involving a variety of driving options.

The options for observational validation of our numerical models are limited.
Interstellar turbulence in general and supersonic turbulence in molecular clouds 
in particular could hardly be uniform and/or isotropic \citep{kaplan.70}. 
There are multiple driving mechanisms of different natures operating on different 
scales in the ISM \citep{norman.96,maclow.04}, so the source function of
turbulence is expected to be broadband. Various observational techniques employed to 
extract information about the scaling properties of turbulence have their own 
limitations, including a finite instrumental resolution, insufficiently large 
data sets, inability to fully access the three-dimensional information without 
additional a priori assumptions, etc. Moreover, complexity of the effective 
equation of state of the ISM and many other physical processes so far ignored 
in the simplified numerical models should also make the comparison of observations 
and simulations uncertain. Nevertheless, it makes sense to compare our results 
with the scaling properties of supersonic turbulence obtained from observations. 

Applying the velocity channel analysis (VCA) technique \citep{lazarian.00,lazarian.06}
to power spectra of integrated intensity maps and single-velocity channel maps of the
Perseus region, \citet{padoan...06} found a velocity power spectrum index $\beta=1.8$
that is reasonably close to our measurement $\beta=1.95$.
The structure function exponents measured for the M1-67 nebula by
\citet{grosdidier....01}, $\zeta_1\approx 0.5$ and $\zeta_2\approx 0.9$, match
quite nicely with our results discussed in Section~\ref{vsf}, $\zeta_1=0.54$ and
$\zeta_2= 0.97$.

Using maps of the $^{13}$CO $J = 1-0$  emission line of the molecular cloud
complexes in Perseus, Taurus, and Rosetta, \citet{padoan...04a} computed the
power spectra of the column density estimated using the LTE\footnote{Local thermodynamic
equilibrium.} method \citep{dickman78}. The slopes of the measured spectra
corrected for temperature and saturation effects on the
$^{13}$CO  $J = 1-0$  line,
$-0.74\pm0.07$, $-0.74\pm0.08$, and $-0.76\pm0.08$, respectively, are notably
shallower than our estimate for the density spectrum power index $-1.07\pm0.01$
(see Section~\ref{dps}). 
The apparent 4~$\sigma$ discrepancy is most probably due to an insufficiently 
high Mach number adopted in our simulations.
 Alternatively, it can be attributed to anisotropies in the molecular cloud 
turbulence, intermittent large-scale driving force acting on the clouds, or 
limitations of the LTE method.

As far as the fractal dimension is concerned, the observational 
measurements for molecular clouds and star-forming regions tend to cluster around 
$D=2.3\pm0.3$ \citep{elmegreen.96,elmegreen.01}. Stochastically variable turbulent 
[WC] winds from ``dustars'' feeding the ISM also demonstrate similar fractal 
dimension $D=2.2-2.3$, and the same morphology of clumps forming and dissipating 
in real time is clearly seen in the outer parts of the wind where the picture 
is not smeared by projection effects \citep{grosdidier....01}.

\section{Conclusions} \label{cncl}
Using large-scale numerical simulations of nonmagnetic highly compressible 
driven turbulence at an rms Mach number of~6, we were able to resolve the 
inertial range scaling and have demonstrated that:

\begin{enumerate}
\item The probability density function of the gas density is perfectly
represented by a lognormal distribution over many decades in probability
as predicted from simple theoretical considerations.

\item Low-order {\em velocity} statistics deviate substantially from Kolmogorov 
laws for incompressible turbulence. Both velocity power spectra and velocity
structure functions show steeper than Kolmogorov slopes, with the
scaling exponents of the third-order velocity structure functions far in 
excess of unity. 

\item The {\em density} power spectrum is instead substantially shallower than
in weakly compressible turbulent flows.

\item The {\em kinetic energy} power spectrum (built on 
$\pmb{w}\equiv\sqrt{\rho}\pmb{u}$) is shallower than Kolmogorov's $5/3$-law.
The power spectra for both the solenoidal and the dilatational parts of $\pmb{w}$
obtained through Helmholtz decomposition ($\pmb{w}=\pmb{w}_S+\pmb{w}_D$) have the
same slopes as the total kinetic energy spectrum pointing to the unique energy 
cascade captured by $E(k)$; their shares in the total energy balance are 68\% and 
32\%, respectively.

\item As should have been expected, the mean volume energy transfer rate in 
compressible turbulent flows, $\rho u^2 u/\ell$, is very close to constant in 
a (statistical) steady state. Our simulations show that the power spectrum of 
$\pmb{v}\equiv\rho^{1/3}\pmb{u}$ scales approximately as $k^{-5/3}$ and the 
third order structure function of $\pmb{v}$ scales linearly with $\ell$ in the inertial 
range.

\item The directly measured mass dimension of the ``fractal'' density distribution
is about 2.4 in the inertial range and 2.0 in the dissipation range. The
geometry of supersonic turbulence is dominated by clustered corrugated
shock fronts.

\end{enumerate}

These results strongly suggest that the Kolmogorov laws originally 
derived for incompressible turbulence would also hold for highly compressible 
flows as soon as they are extended by replacing the pure velocity statistics with
statistics of mixed quantities, such as the density-weighted fluid velocity 
$\pmb{v}\equiv\rho^{1/3}\pmb{u}$. 

Both the scaling properties and the geometry of supersonic turbulence we find 
in our numerical experiments support a phenomenological description of 
intermittent energy cascade in a compressible turbulent fluid
and suggest a reformulation of inertial range intermittency 
models for compressible turbulence in terms of the proposed extension to 
the K41 theory.

The scaling exponents of the velocity diagnostics, the morphology of turbulent
structures, and the fractal dimension of the mass distribution determined in our 
numerical experiments demonstrate good agreement with the corresponding 
observed quantities in supersonically turbulent molecular clouds and line 
radiation-driven winds from carbon-sequence Wolf-Rayet stars.

\acknowledgements
We are grateful to {\AA}ke Nordlund who kindly provided us with his
OpenMP-parallel routines to compute the power spectra.
This research was partially supported by NASA ATP grant NNG05-6601G,
NSF grants AST~05-07768 and AST~06-07675, and NRAC allocation MCA098020S.
We utilized computing resources provided by the San Diego Supercomputer Center
and the National Center for Supercomputer Applications.

\end{document}